\documentclass[useAMS, usenatbib,usegraphicx]{mn2e}
\usepackage{rotating}
\usepackage{amsmath}

\newcommand{\cmt}{{\rm cm}$^{-3}$}

\newcommand{\msun}{M$_\odot$}
\newcommand{\msunpc}{M$_\odot$ pc$^{-2}$}

\newcommand{\twCO}{$^{12}$CO}
\newcommand{\thCO}{$^{13}$CO}

\newcommand{\Halpha}{H$\alpha$}
\newcommand{\sigmol}{$\Sigma_{\rm mol}$}
\newcommand{\sigden}{$\Sigma_{\rm den}$}
\newcommand{\tmol}{$\tau_{\rm dep}^{\rm CO}$}
\newcommand{\tden}{$\tau_{\rm den}$}

\newcommand{\epsden}{$\epsilon_{\rm den}$}
\newcommand{\epsmol}{$\epsilon_{\rm mol}^{\rm CO}$}
\newcommand{\fden}{$f_{\rm den}$}
\newcommand{\fdiff}{$f_{\rm diff}$}
\newcommand{\sigsfr}{$\Sigma_{\rm SFR}$}

\newcommand{\Xunits}{cm$^{-2}$ K$^{-1}$ km$^{-1}$ s}

\newcommand {\apgt} {\ {\raise-.5ex\hbox{$\buildrel>\over\sim$}}\ }
\newcommand {\aplt} {\ {\raise-.5ex\hbox{$\buildrel<\over\sim$}}\ }
\newcommand {\paplt} {\ {(\raise-.5ex\hbox{$\buildrel<\over\sim$}}\ }
\newcommand{\XCO}{$X_{\rm CO}$}
\newcommand{\WCO}{$W_{\rm CO}$}

\newcommand{\Nmol}{$N_{\rm mol}$}

\newcommand{\tsig}{$2\sigma$}

\title[Diffuse CO and the KS relationship] {Interpreting the
  sub-linear Kennicutt-Schmidt relationship: The case for diffuse
  molecular gas} \author[R. Shetty, P. C. Clark, R. S. Klessen]{Rahul
  Shetty$^{1}$, Paul C. Clark$^{1}$, and Ralf S. Klessen$^{1}$
  \\ $^{1}$Universit\"at Heidelberg, Zentrum f\"ur Astronomie,
  Institut f\"ur Theoretische Astrophysik, Albert-Ueberle-Str. 2,
  69120 Heidelberg, Germany}

\begin{document}

\date{Accepted 0000. Received ---; in original form -----}

\pagerange{\pageref{firstpage}--\pageref{lastpage}} \pubyear{2013}
\maketitle

\label{firstpage}
\begin{abstract}

  Recent statistical analysis of two extragalactic observational
  surveys strongly indicate a sublinear Kennicutt-Schmidt (KS)
  relationship between the star formation rate (\sigsfr) and molecular
  gas surface density (\sigmol).  Here, we consider the consequences
  of these results in the context of common assumptions, as well as
  observational support for a linear relationship between \sigsfr\ and
  the surface density of dense gas.  If the CO traced gas depletion
  time (\tmol) is constant, and if CO only traces star forming giant
  molecular clouds (GMCs), then the physical properties of each GMC
  must vary, such as the volume densities or star formation rates.
  Another possibility is that the conversion between CO luminosity and
  \sigmol, the \XCO\ factor, differs from cloud-to-cloud.  A more
  straightforward explanation is that CO permeates the hierarchical
  ISM, including the filaments and lower density regions within which
  GMCs are embedded.  A number of independent observational results
  support this description, with the diffuse gas comprising at least
  30\% of the total molecular content.  The CO bright diffuse gas can
  explain the sublinear KS relationship, and consequently leads to an
  increasing \tmol\ with \sigmol.  If \sigsfr\ linearly correlates
  with the dense gas surface density, a sublinear KS relationship
  indicates that the fraction of diffuse gas \fdiff\ grows with
  \sigmol.  In galaxies where \sigmol\ falls towards the outer disk,
  this description suggests that \fdiff\ also decreases radially.

\end{abstract}


\section{Introduction} \label{introsec}

As observations reveal that stars form predominantly in the molecular
component of the interstellar medium (ISM), the physical conditions of
the H$_2$ gas undoubtedly influence the star formation process.  For
example, the star formation {\it rate} (SFR) is directly related to
the {\it amount} of molecular gas.  This fundamental property is
supported by observations, often through the detection of the CO
($J=1-0$) rotational line as a molecular gas tracer, and some
combination of stellar UV, \Halpha\ from HII regions, and dust thermal
emission in the infrared (IR) due to heating from young stars
\citep[e.g][and references therein]{Kennicutt&Evans12}.  Though the
trend of increasing SFR with higher CO luminosity is unambiguous,
quantifying such correlations, as well as the associated theoretical
interpretations \citep{Maclow&Klessen04, McKee&Ostriker07}, remain a
subject of considerable debate.

One formulation of the correlation between the surface densities of
the star formation rate \sigsfr\ and molecular gas \sigmol\ is the
power-law ``Kennicutt-Schmidt'' \citep[hereafter KS,][]{Schmidt59,
  Kennicutt89} relationship:
\begin{equation}
\Sigma_{\rm SFR} = a \Sigma_{\rm mol}^{N_{\rm mol}}.
\label{KSlaw}
\end{equation}
The surface densities in Equation \ref{KSlaw} are estimated by
employing conversion factors, such as the \XCO\ factor for translating
the CO luminosity to \sigmol\ \citep[see][and references
therein]{Bolatto+13}, and an appropriate factor for estimating
\sigsfr\ from the star formation tracer \citep[see references
in][]{Kennicutt&Evans12}.  The correlation between \sigsfr\ and the
total gas surface density, including the contribution of HI, exhibits
larger scatter \citep{Bigiel+08, Schruba+11}\footnote{The scatter is
  dependent to some extent on the observed scale
  \citep[][]{Onodera+10, Schruba+10, Kim+13, Kruijssen&Longmore14}.} further
indicative of a more direct link between star formation and the
molecular component\footnote{Note, however, that H$_2$ or CO are not
  strictly necessary for star to form, as C$^+$ can also be an
  efficient coolant \citep{Krumholz+11, Krumholz12, Glover&Clark12}.}.

Estimates of the KS parameters $a$ and the index \Nmol\ in Equation 1
range from super-linear \citep[$\sim$1.5, ][]{Kennicutt89, Liu+11,
  Momose+13}, linear \citep{Bigiel+08, Leroy+13}, to sublinear
(\citealt{Shetty+13}, hereafter SKB13, and \citealt{Shetty+14}).
\citet{Kennicutt89, Kennicutt98} measured an index $\approx1.4\, \pm$
0.15 from unresolved observations over entire galactic disks, covering
a range over five orders of magnitude in gas surface density.  These
observations included both normal spirals as well as IR starbursts,
and considered the total HI + H$_2$ gas surface densities.  One
interpretation for a KS index of $\sim$1.5 is that the primary
mechanism in the star formation process is the free fall collapse of
molecular clouds \citep[][]{Elmegreen94, Kennicutt98}.  For a recent
review of explanations of the KS relationship, see \citet{Dobbs+13}.

However, recent resolved extragalactic observations on 100 $-$ 1000 pc
scales demonstrated a tighter molecular KS relationship with lower
indices.  The analysis of the STING and HERACLES surveys advocated for
a linear KS relationship, with significant variations between galaxies
\citep[][]{Leroy+08, Leroy+09a, Leroy+13, Bigiel+08, Schruba+11,
  Rahman+11, Rahman+12}.  The interpretation of a linear KS
relationship is that CO is primarily tracing star forming clouds with
relatively uniform properties, including \sigsfr.  A direct
consequence of this description is that the depletion time of the CO
traced gas is constant and approximately 2 Gyr, both within and
between galaxies.  Evidently, there is no consensus on either the
precise KS parameter estimates, or the associated interpretation.

Two recent statistical analyses of STING and a sub-sample of HERACLES
by SKB13 and \citet{Shetty+14} have indicated that the data actually
favor a sublinear KS relationship for both ensembles, as well as for
most of the individual galaxies.  Those works developed and applied a
Bayesian fitting method that included a treatment of uncertainties,
and provides parameter estimates for each individual galaxy as well as
the population.  SKB13 explained the advantages of a hierarchical
Bayesian method for fitting the KS relationship, and demonstrated its
accuracy over common non-hierarchical methods \citep[see
also][]{Kelly07, Gelman&Hill07, Kruschke11, Gelman+04}.

Other recent efforts favor a sublinear KS relationship as well.  Using
\Halpha\ observations of M51 at 170 pc scales, \citet{Blanc+09} infer
a sublinear KS relationship, \Nmol\ $=$ 0.82 $\pm$ 0.05.
Additionally, \citet{Ford+13} estimate \Nmol\ $\approx$ 0.6 in M31
from observations at a comparable scale.  Finally, \citet{Wilson+12}
find that the ratio of integrated CO ($J=3-2$) to IR luminosity
increases with CO ($J=3-2$) luminosity (see their Fig. 5), suggesting
\Nmol\ $<$ 1.  How can we interpret this emerging evidence for the
sublinear KS relationship?

The standard interpretation is that CO ($J=1-0$) traces ``clouds'' or
``giant molecular clouds'' \citep[GMCs, e.g.][]{Dickman+86, Solomon+87}.
Several theories have attempted to explain the index of the KS
relationship, assuming that GMCs constitute the basic star forming
unit, and that these GMCs are ``virialized''
\citep[e.g.][]{Krumholz&McKee05}.  These assumptions, however, face
difficulty if the CO line is not solely a cloud tracer, but rather
also delineates more diffuse molecular gas distinguishable from the
densest star forming regions.

SKB13 attribute the sublinear KS relationship to the presence of CO
outside of star forming regions, perhaps in a diffuse but pervasive
molecular component.  This description is consistent with a complex,
hierarchical ISM consisting of shells, filaments, low density
ephemeral wisps, besides the well-known high density star forming
clouds.  In this work, we explore this further and consider how and/or
to what extent common assumptions about GMCs can withstand a sublinear
KS relationship.

In the next section, we discuss gas depletion times.  Then, in Section
3 we provide three explanations for a sublinear KS relationship,
including the presence of the diffuse molecular component and its role
in estimating star formation efficiencies.  Subsequently, in Section 4
we describe independent observational investigations revealing diffuse
molecular gas.  We then provide a description of this component in the
hierarchical ISM in Section 5, and additional associated implications
in Section 6.  We conclude with a summary in Section 7.
\section{Implications of a sublinear KS relationship} \label{KSsec}

A corollary of a sublinear KS relationship is that the gas depletion time,
\begin{equation}
\tau_{\rm dep}^{\rm CO}=\Sigma_{\rm mol}/\Sigma_{\rm SFR},
\label{tdepeqn}
\end{equation}
{\it increases} with increasing gas surface density.  As evident in
Figure 2 in \citet{Shetty+14}, a number of galaxies portray this
trend, with only a few galaxies in the STING and \citet{Bigiel+08} surveys
being consistent with a constant \tmol.

\begin{figure}
\includegraphics[width=90mm]{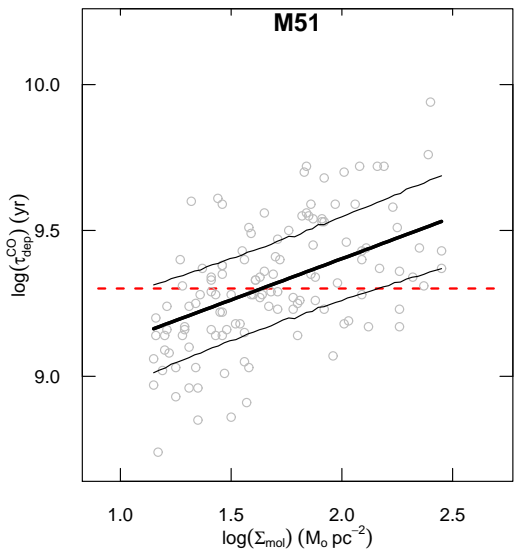}
\caption{Molecular gas depletion times \tmol\ and surface densities
  \sigmol\ in M51.  Grey points are the \tmol\ and \sigmol\ estimated
  from the observations \citep[and publically available
  in][]{Bigiel+10c}.  Thick and thin lines mark the median and \tsig\
  range of the linear regression from SKB13.  Dashed line marks a
  constant \tmol\ = 2 Gyr.}
\label{M51tdep}
\end{figure}

Figure \ref{M51tdep} shows estimates of the depletion time and surface
density of M51, estimated from BIMA SONG CO observations
\citep{Helfer+03}, FUV Nearby Galaxy Survey
\citep[NGS;][]{GildePaz+07} and the Spitzer SINGS survey
\citep[][]{Kennicutt+03}\footnote{The \sigsfr\ and \sigmol\ estimated
  from these datasets are publically available in \citet{Bigiel+10c}}.
The points show \sigmol, which is the product of the observed
luminosities and the standard \XCO\ factor, and \tmol\ obtained
through Equation (\ref{tdepeqn}).  The dashed line marks a constant
\tmol\ = 2 Gyr.  Clearly, \tmol\ increases with \sigmol.
The thick solid line in Figure \ref{M51tdep} is the predicted trend
for \Nmol\ = 0.72, which is the most likely value from the
hierarchical Bayesian fit from SKB13.  The thin solid lines indicate
the range of plausible KS fits at 95\% confidence from SKB13.  Table
\ref{tdeptab} lists the 2.5\%, 50\%, and 97.5\% quantiles of \tmol\ at
five different values of \sigmol.  The most likely value of \tmol\
increases from 1.7 Gyr at 25 \msunpc\ to 3 Gyr at 200 \msunpc.

\begin{table}
 \centering
 \begin{minipage}{140mm}
  \caption{Estimated Depletion Times in M51}
  \begin{tabular}{cccc}
  \hline
  \hline
\sigmol & \tmol\ (2.5\%) & \tmol (50\%) & \tmol (97.5\%)  \\
(\msunpc) & (Gyr) & (Gyr) & (Gyr)  \\
\hline
25   & 1.2 & 1.7 & 2.4 \\
50   & 1.5 & 2.1 & 2.9 \\
100  & 1.8 & 2.5 & 3.5 \\
150  & 2.0 & 2.8 & 4.0 \\
200  & 2.2 & 3.1 & 4.4 \\
\hline
\end{tabular}
\label{tdeptab}
\end{minipage}
\end{table}

The inverse of \tmol\ is the rate of star formation per unit surface
density, often referred to as the star formation ``efficiency''
\epsmol\ per unit time.  This rate corresponds to the fraction of CO
traced gas converted into stars over some timescale.  Figure
\ref{M51tdep} indicates that there are some fundamental differences in
the star formation properties as the gas surface densities varies.

\section{Origins of a sub-linear KS Relationship} \label{expKSsec}

In this section, we consider three possible explanations for the
inferred sublinear KS relationship.  The first two are systematic
effects that produce the observed trends when the underlying KS
relationship is linear.  In these subsections, we explore the
possibility that the inferred variable and sublinear relationship
arises due to variations in cloud properties and the \XCO\ factor.
However, given the wide range in KS slopes, these scenarios are
unlikely.  The third interpretation is that CO is also tracing
significant amounts of diffuse molecular gas.

\subsection{Clouds with different properties?} \label{diffclouds}

A popular explanation for a linear KS relationship is that the
observed CO luminosity is directly proportional to the number of
star-forming clouds or GMCs, with all clouds having similar
properties, such as the volume density, the efficiency of the cloud,
and the star formation rate.  In observations with 100 $-$ 1000 pc
resolutions, the individual clouds are not resolved, but rather their
CO flux is dispersed throughout the beam.  Under this assumption,
regions with more clouds emit more CO, in proportion to the number of
clouds.  Extragalactic CO observations, therefore, are simply
``counting clouds''.

However, the sublinear KS relationship suggest that the clouds do not
have the same properties, so there is no one-to-one correspondence
between the CO luminosity and the number of clouds in the beam.  Two
likely possibilities are that the star formation efficiencies vary,
and/or that the volume densities of the clouds are not constant.

For instance, if the properties of star-forming clouds are highly
sensitive to the effects of feedback, then the efficiency would depend
on the evolutionary state of the cloud.  Younger clouds, which have
yet to form (many) stars, are unaffected by feedback.  As the cloud
evolves, feedback processes from within begin to dramatically alter
the cloud, until it is eventually destroyed.  Supporting this
description, \citet{Murray11} and \citet{Battisti&Heyer14} infer a
wide range of efficiencies (per free-fall time), and dense gas
fractions, respectively, spanning over an order of magnitude
\citep[see also][]{MurrayRahman10}.  These results suggest that
individual clouds could have a time-dependent efficiencies.  The
systematic trend of a decreasing \epsmol\ with increasing \sigmol\
(Fig. \ref{M51tdep}) could be indicative of a decreasing efficiency
with increasing GMC mass, if CO is indeed only tracing star-forming
clouds.

Another closely related possibility is that the volume density differs
between clouds.  Such differences may also lead to a variable
efficiencies.  Quantifying the densities of clouds requires knowledge
of the cloud masses and sizes.  As the clouds are unresolved to begin
with, it is not possible to do so with the CO observations alone.

\subsection{Variations in the \XCO\ factor?} \label{diffXco}

\begin{figure}
\includegraphics*[width=90mm]{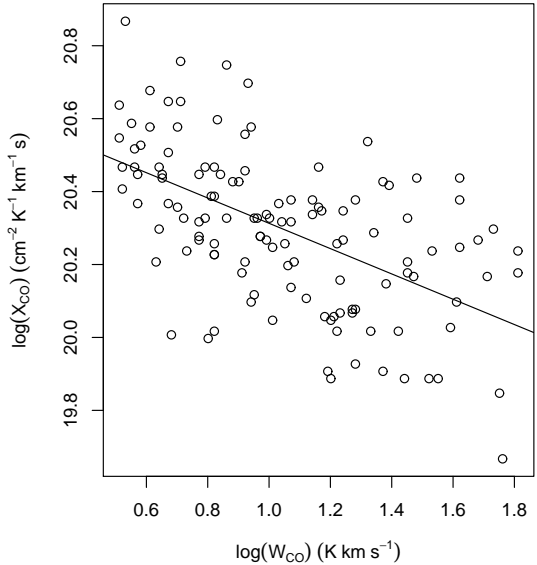}
\caption{The variation of the conversion factor \XCO\ with CO
  intensity \WCO\ required for \tmol\ = 2 Gyr.  The line shows the
  linear regression result, which has a slope of $-$0.3, [consistent
  with \Nmol\ = 0.7, see Equations (\ref{sigL2})-(\ref{Xbeta})].}
\label{varXco}
\end{figure}

The measured values of \sigmol\ depend on the assumptions in
translating the CO brightness to gas densities.  Most studies to date
employ the standard Galactic value \XCO\ = 2$\times10^{20}$ \Xunits\
to convert the observed intensity \WCO\ to H$_2$ column densities.  As
\sigmol\ is a surface density,
\begin{equation}
\Sigma_{\rm mol} \propto X_{\rm CO} W_{\rm CO}.
\label{Xfaceqn}
\end{equation}
For an assumed constant \XCO, Equation 1 simply states:
\begin{equation}
\Sigma_{\rm SFR} \propto W_{\rm CO}^{N_{\rm mol}}.
\label{sigL}
\end{equation}
If $\Sigma_{\rm SFR} \propto \Sigma_{\rm mol}$,
\begin{equation}
\Sigma_{\rm SFR} \propto X_{\rm CO} W_{\rm CO}.
\label{sigL2}
\end{equation}
If the estimated value of \Nmol\ is not unity, \XCO\ would have to
vary with \WCO.  Let us consider the relation:
\begin{equation}
X_{\rm CO} \propto  W_{\rm CO}^\beta.
\label{sigL3}
\end{equation}
Together, Equations (\ref{sigL}) - (\ref{sigL3}) require that
\begin{equation}
\beta = N_{\rm mol} - 1.
\label{Xbeta}
\end{equation}
If the \tmol\ = 2 Gyr, then we could solve
\begin{equation}
X_{\rm CO} \propto \frac{(2\times10^9\, {\rm yr}^{-1}) \, \Sigma_{\rm SFR}}{W_{\rm CO}}.
\label{Xabs}
\end{equation}
Figure \ref{varXco} shows the \XCO\ $-$ \WCO\ relationship required to
produce a contant \tmol\ = 2 Gyr for M51.  We find that \XCO\ varies
in the range 19.8 \aplt\ log(\XCO/[\Xunits]) \aplt\ 20.8 \Xunits.  The
line shows a linear regression fit to the data, with slope
$\beta=-0.3$, which suggests a steeper \XCO\ $-$ \WCO\ dependence
compared to results relying on the assumption of virial equilibrium
\citep{Solomon+87}.  This slope can be directly estimated from
Equation \ref{Xbeta} with \Nmol\ = 0.7, which is the most likely
estimated KS slope for M51 (SKB13).

As indicated by Equation (\ref{Xbeta}), in order for \tmol\ to be
constant, there should be an inverse correlation between \XCO\ and
\WCO\ if \Nmol\ $<$ 1.  The precise value of $\beta$ depends on the
magnitude of the sublinearity, and therefore the properties of the
individual galaxies.  Indeed, theoretical and observational efforts
have clearly demonstrated that \XCO\ differs in environments with a
range of metallicities, radiation fields, and/or turbulence levels
\citep[e.g.][]{vanDishoeck&Black88, Maloney&Black88, Rubio+04,
  Glover&Maclow11, Leroy+11, Ostriker&Shetty11, Shetty+11a,
  Narayanan+11, Narayanan+12, Sandstrom+13, Lee+14}.  However, many
galaxies in the STING and HERACLES sub-samples investigated by SKB13
and \citet{Shetty+14} are local universe star-forming galaxies, and
are therefore believed to have similar (solar) metallicities and
turbulence levels.  In order to recover an ``universal'' linear KS
relationship, the \XCO\ factor would have to conspire with \WCO\ to
offset the variability in \Nmol\ between individual galaxies.  It is
therefore unlikely that the \XCO\ significantly changes between these
galaxies, though a thorough assessment is necessary before ruling this
possibility out.  We further discuss the \XCO\ factor in Section 6.

\subsection{Existence of a substantial diffuse molecular component?} \label{diffgas}

The third possibility for an increasing \tmol\ with rising \sigmol\ is
that diffuse molecular gas is contributing to the observed CO
luminosity.  The primary difference between this scenario and the
previous two is that CO is not necessarily solely a tracer of star
forming clouds.  Figure \ref{M51tdep} indicates that the diffuse
component contributes more strongly towards \WCO\ with increasing
\sigmol, relative to the usual contribution from star forming clouds
or GMCs.  Here, we consider a simple equilibrium model of the ISM
consisting of two generic phases, a ``dense'' and a ``diffuse''
component.

In equilibrium the rate at which gas turns from the diffuse into the
dense star-forming phase is constant in time.  In the multi-phase ISM,
this rate includes the time required for core formation within GMCs,
and subsequently star formation within the cores.  This translates
into constant cloud, core, and star formation efficiencies.  On the
large scales appropriate for extragalactic observations, recent
theoretical efforts expect such an equilibrium condition to hold
\citep[e.g.][]{Elmegreen02, Shetty&Ostriker08, Ostriker+10,
  Ostriker&Shetty11, Kim+11, Dobbs+11b, Hopkins+11}.  This equilibrium
implies a linear relationship between the star formation rate and the
surface density of the most dense gas, \sigden, supported by recent
observational efforts \citep[e.g.][]{Heiderman+10, Lada+12}.
Accordingly, the interpretation of a constant gas depletion time,
invoked for \Nmol\ $\approx$ 1, may be applicable not to all CO
$J=1-0$ traced gas, but rather some other higher density component.
This dense gas may be traced by higher level CO transitions, other
dense molecular gas tracers such as HCN (with critical densities
$\sim10^6$\cmt), or gas with extinctions above a threshold
\citep[e.g.][]{Gao&Solomon04, Wu+05, Lada+10}.  Accordingly, we
hereafter consider a model where the star formation rate scales
linearly with the surface density of dense gas, as previously
discussed by \citet{Lada+10} and \citet{Heiderman+10}.

We assume the presence of a high density component that scales
linearly with \sigsfr, but do not specify the threshold density or the
tracer of this component, though we discuss the viability of CO as
such a dense gas tracer in the next Section.  We will refer to the
surface density of this gas with \sigden, so that
\begin{equation}
\Sigma_{\rm SFR} = \epsilon_{\rm den} \Sigma_{\rm den},
\label{KSdense}
\end{equation}
where \epsden\ is the efficiency per unit time.  The inverse of the
efficiency is the dense gas depletion time
\begin{equation}
\tau_{\rm den} = \Sigma_{\rm den}/\Sigma_{\rm SFR}.
\label{taudense}
\end{equation}

In this simple equilibrium model of star formation, \tden\ is
constant.  We can now define the dense gas fraction \fden:
\begin{equation}
f_{\rm den} = \Sigma_{\rm den}/\Sigma_{\rm
  mol} = \tau_{\rm den} / \tau_{\rm dep}^{\rm CO},
\label{feqn}
\end{equation}
using the definition of \tmol\ in Equation (\ref{tdepeqn}).  The
fraction of diffuse molecular gas, i.e. the component not directly
forming stars, is \fdiff = 1 $-$ \fden.

By construction, \tden\ is constant everywhere within a given galaxy.
\tmol, on the other hand, is only constant if and only if \Nmol\ = 1:
\[
    f_{\rm den} =
\begin{cases}
    {\rm constant},& \text{if } N_{\rm mol}=1\\
    f_{\rm den}(\Sigma_{\rm mol}), & \text{if } N_{\rm mol}\neq 1
\label{fconds}
\end{cases}
\]
Under this framework, \fden\ depends on the local properties of the
ISM, such as \sigsfr\ and \sigmol, unlike the global galactic
properties such as \Nmol\ (and by assumption \tden\ and thereby
\epsden).

\begin{figure}
\includegraphics*[width=90mm]{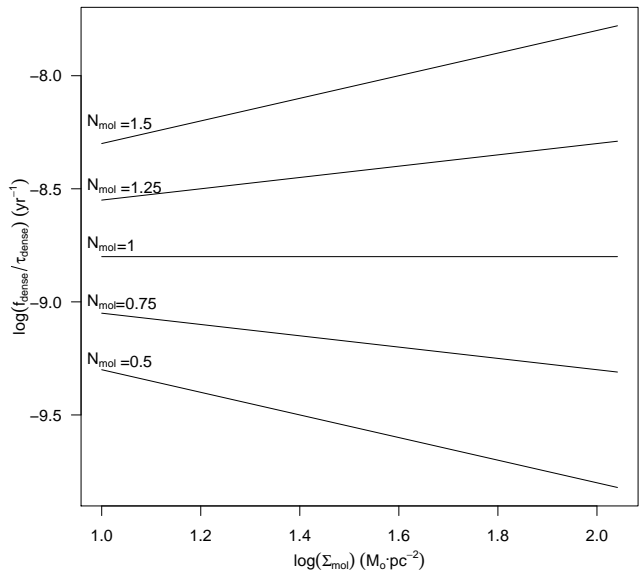}
\caption{The dense gas fraction \fden\ as a function of \sigmol, for
  different KS slopes.  The ordinate provides the ratio \fden/\tden,
  where \tden\ is the depletion time of the dense gas (see Equation
  \ref{taudense}).}
\label{fvss}
\end{figure}

Figure \ref{fvss} shows how the ratio \fden/\tden\ varies with
\sigmol.  Note that \fden/\tden\ is simply the inverse of \tmol\ (see
Eqns. \ref{tdepeqn} and \ref{feqn}) given our assumption that \tden\
is constant.  We show the ratio \fden/\tden, rather than just \fden,
as the absolute value of \fden\ depends on \tden.  Figure \ref{fvss}
indicates that \fden\ decreases by half an order of magnitude between
10 $-$ 100 \msunpc\ for \Nmol\ = 0.5 \citep[e.g. for NGC
772,][]{Shetty+14}.  In M51, \Nmol\ = 0.75, so \fden\ decreases by a
factor of 2 between 10 $-$ 100 \msunpc, since \tden\ is constant, and
\tmol\ varies from $\approx$ 1.7 to 2.5 Gyr.  Now, if free-fall
collapse at typical GMC densities $\sim$100 \cmt\ govern \tden, then
\fden\ is only a fraction of a percent (0.1$-$0.2\% in M51) of the
total molecular content.  Alternatively, if a galaxy has \Nmol $>$ 1,
then \fden\ increases and \fdiff\ decreases with \sigmol.  When \Nmol\
= 1, \fden\ and \fdiff\ are constant, and are therefore global
properties of the galaxy.

We have shown that in a simple equilibrium description of the ISM, the
variation of \fden\ within an individual galaxy depends on \Nmol.
Unless \Nmol\ = 1, then \fden\ and \fdiff\ varies with \sigmol.  As
most galaxies evince a sublinear KS relationship, \fden\ (\fdiff)
decreases (increases) with increasing \sigmol.  In galaxy disks
\sigmol\ decreases with radius, so \fdiff\ consequently also falls
radially.  In the next section, we review previous observations
detecting the diffuse molecular component.

\section{Observational evidence for diffuse molecular gas} \label{diffgassevid}


Perhaps the first proof of CO emission emerging from regions other
than from star-forming clouds was the detection of ``high latitude
clouds'' (HLCs) in the solar neighborhood \citep{Blitz+84}.  Larger
surveys confirmed the ubiquity of such diffuse HLCs
\citep[][]{Magnani+00, Onishi+01}.  These clouds differ from the
conventional star-forming clouds in that they have smaller radii, R
\aplt 10 pc, and lower masses ($M$ \aplt 100 \msun).  Although these
clouds are very low mass compared to GMCs, they are numerous, and
hence are likely to contribute significantly towards the observed CO
intensity in extragalactic observations.

Within galactic disks, diffuse emission may contribute substantially
to the observed emission.  The identification of molecular clouds from
spectral line data customarily involve decomposition techniques.
Namely, contiguous voxels in a position-position-velocity cube above a
chosen threshold are identified as a cloud or GMC, which may not
necessarily correspond to real features
\citep[][]{Ballesteros-Paredes&MacLow02, Gammie+03, Shetty+10,
  Beaumont+13}.  Nevertheless, integrated intensities represents an
estimate of the total content.  \citet{Rosolowsky+07} found that 40\%
$-$ 80\% of the CO luminosity, depending on the radius, occurs away
from the massive (10$^5$ \msun) clouds.  In the MW,
\citet{SolomonRivolo89} find that $\sim$60\% of the CO luminosity may
emerge from clouds with masses smaller than 10$^4$ \msun.  The
remaining emission originates from lower mass clouds, or simply from a
more extended, diffuse component.  Additionally, recent combined
interferometric and single-dish CO observations of M51 have unveiled a
thick disk of diffuse molecular gas \citep[][]{Schinnerer+13, Pety+13,
  Hughes+13}.  This diffuse component accounts for nearly 50\% of the
detected CO luminosity\footnote{Given the limited physical resolution
  of these extra-galactic observations, these diffuse gas fractions
  are likely lower limits.}.  This large contribution from the diffuse
component in M51 certainly affects the derived sublinear KS
relationship, with \Nmol\ $\approx$ 0.7 $-$ 0.8
\citep[][SKB13]{Blanc+09}.  \citet{Wilson&Walker94} measure a higher
\twCO\ to \thCO\ ratio from single-dish observations M33, compared to
the ratio inferred from interferometric observations of an individual
cloud in M33.  \citet{Wilson&Walker94} attribute the larger ratio from
the single-dish observations to the presence of diffuse clouds
(i.e. not GMCs), as found in the MW by \citet{Polk+88}.  They place a
lower limit on the amount of this diffuse emission at 30\% of the
total CO intensity \citep[see also][]{Burgh+07, Liszt+10}.

If CO is prevading the entire galaxy, the thicknesses measured in the
atomic and molecular components should not differ.  In fact, the
recent observational findings by \citet{Caldu-Primo+13} of very
similar HI and CO linewidths from extragalactic observations suggests
that both CO and HI are tracing the full vertical extent of the ISM,
rather than an ordered medium with disparate molecular clouds embedded
in a dominant diffuse atomic medium.  This argues in favor of a
diffuse but volume filling molecular component.

\section{CO in the hierarchical ISM} \label{diffgastheory}

These observations attest to the presence of diffuse CO, but do not
reveal how this component is organized.  The structure of the ISM is
known to be hierarchical, such that there are dense features embedded
within lower density regions on all mass or length scales.  For
example, the direct precursors to stars are the densest ``cores'',
which may form at the intersection of lower density filaments or
clumps, which themselves are embedded within GMCs.  GMCs may further
be situated in some larger scale structure.  Observations suggest that
the ISM is self-similar, in the sense that the statistical properties
of the hierachical structure is similar on all scales.  For an
in-depth review of cloud formation in a hierachical ISM, see
\citet[and references therein]{Elmegreen93PPL, Elmegreen13}.

Turbulent motions are observed on all scales beyond the densest cores,
and likely plays a dominant role in sculpting the self-similar
hierachical ISM \citep{Falgarone+92, Elmegreen02}.  Turbulence
contributes both to the formation and destruction of high density
features \citep[e.g.][]{Elmegreen93b, Elmegreen&Scalo04,
  Scalo&Elmegreen04, Maclow&Klessen04}.  It may cause the compression
of dense regions to eventually assemble into a star-forming cloud or
core.  Alternatively, turbulent rarefaction waves may prevent the
collapse of pre-existing structures.

Due in part to the complexity of this turbulent, hierarchical ISM, it
may be too simplistic too identify and/or assign any observed feature
as a cloud \citep{Scalo90}.  One classification scheme divides
``bound'' clouds as those that collapse to form stars, and ``diffuse''
clouds as those that dissipate before star can form.
\citet{Elmegreen93} considers both bound and diffuse clouds, and
argues that the dynamic state of a cloud depends on the relative
effects of the internal pressure to self-gravity.  The chemical state
of the cloud also depends on the local radiation field.  Following
this description, transient events due to star formation, density or
turbulent waves, or passing stars may cause rapid fluctions in the
chemical state of a given region of the ISM.  \citet{Pringle+01}
suggest that observed star forming clouds are simply the peaks of a
hierachical, molecular ISM.  Thus, regardless of its status as a
distinct cloud, the chemical state of a patch of the ISM may not be
directly correlated with its ability to form stars \citep[see
also][]{Glover&Clark12}.

\begin{figure}
\centerline{\includegraphics*[width=70mm]{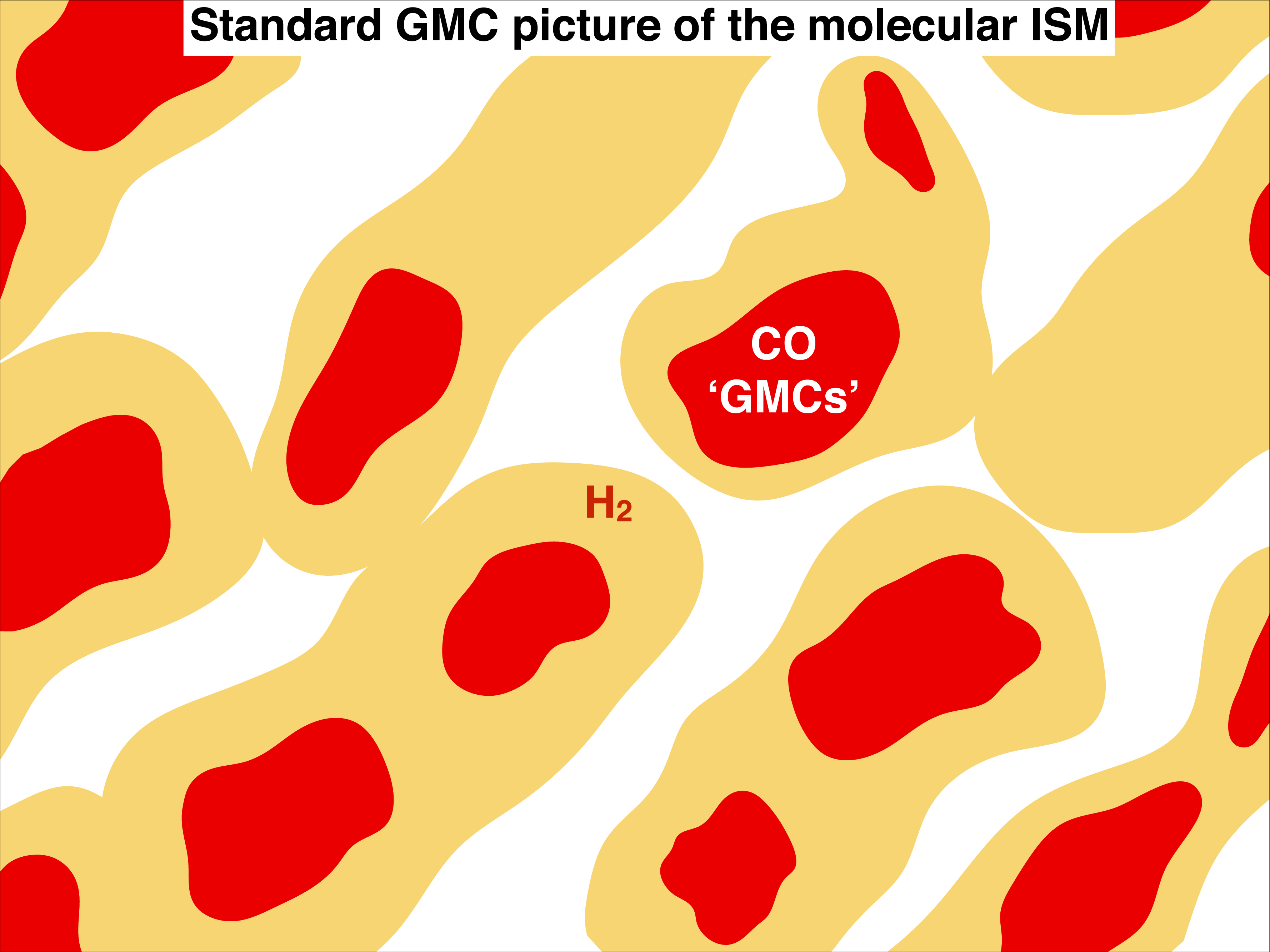}}
\centerline{\includegraphics*[width=70mm]{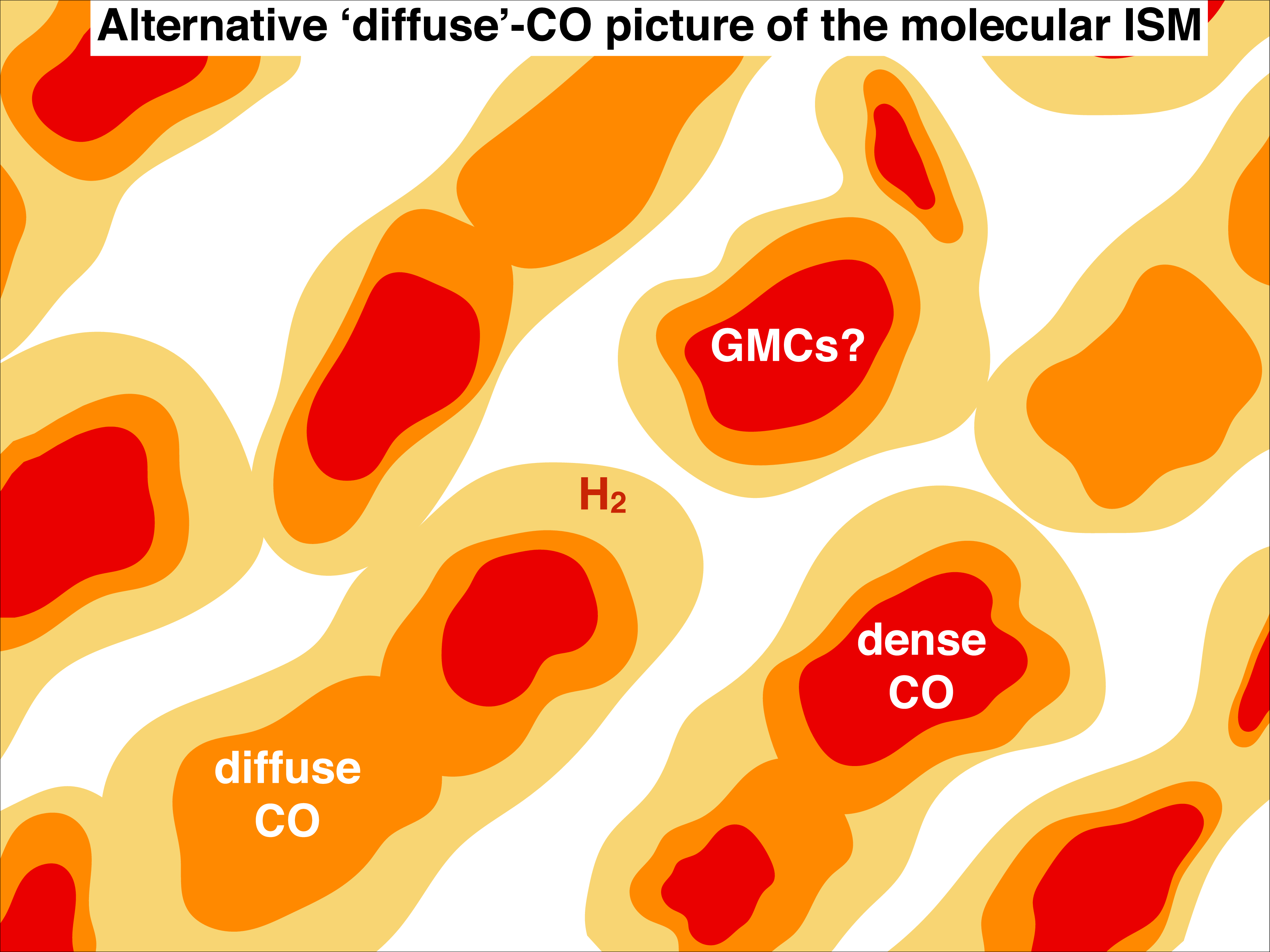}}
\caption{Top: ISM where CO solely traces star-forming GMCs.  Bottom:
  CO (orange and red) is more pervasive than star forming clouds (red)
  in the molecular ISM.}
\label{gmccartoon}
\end{figure}

Figure \ref{gmccartoon} shows a diagram of the hierarchical ISM,
differentiating the scenario where CO only traces GMCs, and that where
CO exists more pervasively.  In the standard framework, CO only traces
GMCs, which are the exclusive sites of star formation.  A sub-linear
KS relationship modifies this description to include CO in more
diffuse regions, as depicted in the bottom panel of Figure
\ref{gmccartoon}.  Future observational analyses are needed to
quantify the relative amounts of dense and diffuse molecular gas, as
well as the physical properties of these phases.

\section{Further implications of diffuse CO emission}

Both our work here, along with that of \citet{Wilson&Walker94},
\citet{Rosolowsky+07}, and \citet{Pety+13} suggest that a large
fraction of the CO ($J=1-0$) intensity should be attributed to the
diffuse molecular component, consisting of at least 30\% of the total
molecular mass from extragalactic observations.  The kinematic
distance ambiguity presents an additional important challenge in
Galactic observations, though the contribution from diffuse molecular
gas may be constrained statistically.  Having \thCO\ observations may
help in distinguishing the diffuse and dense phases.  As the CO
($J=1-0$) line is optically thick, it may be impossible to
unambiguously identify the star-forming clouds from the intercloud
medium.

Another possibility for distinguishing between the phases is through
higher level CO transitions.  These lines require higher temperatures
and/or densities for excitation.  \citet{Krumholz&Thompson07} and
\citet{Narayanan+08} suggest that the inferred KS index of higher
level CO transitions should be less than the derived \Nmol\ from the
($J=1-0$) line.  In their proposition, the ($J=1-0$) should recover
the underlying star formation law.  This occurs because this line
is thermalized almost everywhere, due to its low critical density, and
thus faithfully follows the intrinsic star formation law.  However,
the upper level transitions have higher critical densities, and thus
only trace a small fraction of the molecular content.  They suggest
that this leads to a shallower inferred KS slope.  It will be
interesting to compare the inferred relationships using different
tracers, which should be possible with ALMA \citep[as well as the JCMT
NGLS survey,][]{Wilson+09}.

Finally, we note that since CO cannot be used as a reliable cloud
tracer, the parameters of the integrated spectral line might neither
provide accurate information about cloud dynamics, nor about the \XCO\
factor.  Nevertheless, cloud properties such as mass, velocity
dispersion, and virial state are often estimated from the CO ($J=1-0$)
line.  Results from these analyses are certainly affected by the
assumption that CO neatly traces clouds, as well as the uncertainties
raised above \citep[see also][]{Pringle+01}.  As elucidated by
\citet{Maloney90}, the standard \XCO\ factor can be derived
analytically based on the assumption of virial equilibrium, since both
the mass and luminosity are set by the CO linewidth \citep[see
also][]{Shetty+11b, Wall07, Narayanan&Hopkins13}.  Consequently, the
presence of diffuse emission complicates any investigation of cloud
structure and dynamics solely from CO observations.

\section{Summary} \label{summarysec}

We have examined possible physical interpretations of the sublinear KS
relationship \citep[SKB13][]{Blanc+09, Ford+13, Shetty+14}.  In
Section \ref{diffclouds}, we indicated that if CO uniquely traces
star-forming clouds, then cloud properties such as the volume density
or star formation efficiency must differ between clouds.  Similarly,
we discussed variations in the \XCO\ factor in Section \ref{diffXco}.

The third possibility, considered in Section \ref{diffgas}, is the
presence of substantial amounts of diffuse molecular gas which also
contributes towards the total CO luminosity.  As the star formation
rate is expected to be linearly correlated with dense gas, then the
resulting KS index \Nmol\ depends on \fdiff.  Galaxies with \Nmol\ $<$
1, such as M51, have \fdiff\ (dense gas fractions \fden) increasing
(decreasing) with \sigmol.  Accordingly, we expect \fdiff\ to drop at
larger radii where \sigmol\ decreases.  Since the KS relationships are
different between galaxies, the \fdiff $-$ \sigmol\ correlations also
correspondingly vary.  Indeed, observations including higher level CO
transitions as well as \thCO\ indicate the presence of substantial
amounts of CO in a diffuse component consisting of \apgt\ 30\%.  This
phase may exist in the form of low mass \paplt 10$^4$ \msun) clouds,
or as a hierarchical and pervasive medium.

Quantifying the amounts of gas in the various phases is necessary for
understanding the timescales associated with star formation.  We
suggest that the sublinearity in the KS relationship is directly due
to the dominant contribution of diffuse CO gas, with large \fdiff.
This results in a long CO depletion time, which in the case of M51,
varies from $\sim$1.7 to \aplt 3 Gyr for 25 \aplt\ \sigmol\ \aplt\ 200
\msunpc.  If collapse only occurs in dense gas at a constant
timescale, then for galaxies such as M51 with \Nmol\ $\approx$ 0.7 the
fraction of CO traced gas currently forming stars is only of order
0.1\% or less.  Future observational analysis, including other ISM
tracers, should further reveal the role of the different phases,
including the timescales and efficiencies in the phase transitions
towards the formation of stars.

\section*{Acknowledgements}

We are very grateful to B. Kelly for his role in our statistical
analysis of the STING and HERACLES sub-samples.  We appreciate
comments on the draft by B. Elmegreen, M. Heyer, M. Y. Lee, and
J. Roman-Duval.  We also thank our colleagues F. Bigiel, A. Bolatto,
C. Dullemond, S. Glover, A. Goodman, C. Hayward, A. Hughes,
L. Konstandin, A. Leroy, S. Meidt, E. Ostriker, J. Pety, N. Rahman,
E. Schinnerer, R. Smith, \& L. Sz\H ucs for extensive collaborations and
discussions over the years which substantially contributed to our
understanding of CO and star formation.  RS, PCC, RSK, and acknowledge
support from the Deutsche Forschungsgemeinschaft (DFG) via the SFB 881
(subproject B1, B2 and B5) ``The Milky Way System,'' and the SPP
(priority program) 1573.  RSK also acknowledges support from the
European Research Council under the European Community’s Seventh
Framework Programme (FP7/2007-2013) via the ERC Advanced Grant {\em
  STARLIGHT} (project number 339177).

\bibliography{citations}

\begin{thebibliography}{99}
\expandafter\ifx\csname natexlab\endcsname\relax\def\natexlab#1{#1}\fi

\bibitem[{{Ballesteros-Paredes} \& {Mac
  Low}(2002)}]{Ballesteros-Paredes&MacLow02}
{Ballesteros-Paredes} J., {Mac Low} M.-M., 2002, \apj, 570, 734

\bibitem[{{Battisti} \& {Heyer}(2014)}]{Battisti&Heyer14}
{Battisti} A.~J., {Heyer} M.~H., 2014, \apj, 780, 173

\bibitem[{{Beaumont} {et~al}\mbox{.}(2013){Beaumont}, {Offner}, {Shetty},
  {Glover}, \& {Goodman}}]{Beaumont+13}
{Beaumont} C.~N., {Offner} S.~S.~R., {Shetty} R., {Glover} S.~C.~O., {Goodman}
  A.~A., 2013, \apj, 777, 173

\bibitem[{{Bigiel} {et~al}\mbox{.}(2010){Bigiel}, {Leroy}, {Walter}, {Blitz},
  {Brinks}, {de Blok}, \& {Madore}}]{Bigiel+10c}
{Bigiel} F., {Leroy} A., {Walter} F., {Blitz} L., {Brinks} E., {de Blok}
  W.~J.~G., {Madore} B., 2010, \aj, 140, 1194

\bibitem[{{Bigiel} {et~al}\mbox{.}(2008){Bigiel}, {Leroy}, {Walter}, {Brinks},
  {de Blok}, {Madore}, \& {Thornley}}]{Bigiel+08}
{Bigiel} F., {Leroy} A., {Walter} F., {Brinks} E., {de Blok} W.~J.~G., {Madore}
  B., {Thornley} M.~D., 2008, \aj, 136, 2846

\bibitem[{{Blanc} {et~al}\mbox{.}(2009){Blanc}, {Heiderman}, {Gebhardt},
  {Evans}, \& {Adams}}]{Blanc+09}
{Blanc} G.~A., {Heiderman} A., {Gebhardt} K., {Evans}, II N.~J., {Adams} J.,
  2009, \apj, 704, 842

\bibitem[{{Blitz} {et~al}\mbox{.}(1984){Blitz}, {Magnani}, \&
  {Mundy}}]{Blitz+84}
{Blitz} L., {Magnani} L., {Mundy} L., 1984, \apjl, 282, L9

\bibitem[{{Bolatto} {et~al}\mbox{.}(2013){Bolatto}, {Wolfire}, \&
  {Leroy}}]{Bolatto+13}
{Bolatto} A.~D., {Wolfire} M., {Leroy} A.~K., 2013, \araa, 51, 207

\bibitem[{{Burgh} {et~al}\mbox{.}(2007){Burgh}, {France}, \&
  {McCandliss}}]{Burgh+07}
{Burgh} E.~B., {France} K., {McCandliss} S.~R., 2007, \apj, 658, 446

\bibitem[{{Cald{\'u}-Primo} {et~al}\mbox{.}(2013){Cald{\'u}-Primo}, {Schruba},
  {Walter}, {Leroy}, {Sandstrom}, {de Blok}, {Ianjamasimanana}, \&
  {Mogotsi}}]{Caldu-Primo+13}
{Cald{\'u}-Primo} A., {Schruba} A., {Walter} F., {Leroy} A., {Sandstrom} K.,
  {de Blok} W.~J.~G., {Ianjamasimanana} R., {Mogotsi} K.~M., 2013, \aj, 146,
  150

\bibitem[{{Dickman} {et~al}\mbox{.}(1986){Dickman}, {Snell}, \&
  {Schloerb}}]{Dickman+86}
{Dickman} R.~L., {Snell} R.~L., {Schloerb} F.~P., 1986, \apj, 309, 326

\bibitem[{{Dobbs} {et~al}\mbox{.}(2011){Dobbs}, {Burkert}, \&
  {Pringle}}]{Dobbs+11b}
{Dobbs} C.~L., {Burkert} A., {Pringle} J.~E., 2011, \mnras, 417, 1318

\bibitem[{{Dobbs} {et~al}\mbox{.}(2013){Dobbs}, {Krumholz},
  {Ballesteros-Paredes}, {Bolatto}, {Fukui}, {Heyer}, {Mac Low}, {Ostriker}, \&
  {V{\'a}zquez-Semadeni}}]{Dobbs+13}
{Dobbs} C.~L. {et~al.}, 2013, ArXiv e-prints

\bibitem[{{Elmegreen}(1993{\natexlab{a}})}]{Elmegreen93PPL}
{Elmegreen} B.~G., 1993{\natexlab{a}}, in Protostars and Planets III, {Levy}
  E.~H., {Lunine} J.~I., eds., pp. 97--124

\bibitem[{{Elmegreen}(1993{\natexlab{b}})}]{Elmegreen93b}
{Elmegreen} B.~G., 1993{\natexlab{b}}, \apjl, 419, L29

\bibitem[{{Elmegreen}(1993{\natexlab{c}})}]{Elmegreen93}
{Elmegreen} B.~G., 1993{\natexlab{c}}, \apj, 411, 170

\bibitem[{{Elmegreen}(1994)}]{Elmegreen94}
{Elmegreen} B.~G., 1994, \apjl, 425, L73

\bibitem[{{Elmegreen}(2002)}]{Elmegreen02}
{Elmegreen} B.~G., 2002, \apj, 577, 206

\bibitem[{{Elmegreen}(2013)}]{Elmegreen13}
{Elmegreen} B.~G., 2013, in IAU Symposium, Vol. 292, IAU Symposium, {Wong} T.,
  {Ott} J., eds., pp. 35--38

\bibitem[{{Elmegreen} \& {Scalo}(2004)}]{Elmegreen&Scalo04}
{Elmegreen} B.~G., {Scalo} J., 2004, \araa, 42, 211

\bibitem[{{Falgarone} {et~al}\mbox{.}(1992){Falgarone}, {Puget}, \&
  {Perault}}]{Falgarone+92}
{Falgarone} E., {Puget} J.-L., {Perault} M., 1992, \aap, 257, 715

\bibitem[{{Ford} {et~al}\mbox{.}(2013){Ford}, {Gear}, {Smith}, {Eales}, {Baes},
  {Bendo}, {Boquien}, {Boselli}, {Cooray}, {De Looze}, {Fritz}, {Gentile},
  {Gomez}, {Gordon}, {Kirk}, {Lebouteiller}, {O'Halloran}, {Spinoglio},
  {Verstappen}, \& {Wilson}}]{Ford+13}
{Ford} G.~P. {et~al.}, 2013, \apj, 769, 55

\bibitem[{{Gammie} {et~al}\mbox{.}(2003){Gammie}, {Lin}, {Stone}, \&
  {Ostriker}}]{Gammie+03}
{Gammie} C.~F., {Lin} Y.-T., {Stone} J.~M., {Ostriker} E.~C., 2003, \apj, 592,
  203

\bibitem[{{Gao} \& {Solomon}(2004)}]{Gao&Solomon04}
{Gao} Y., {Solomon} P.~M., 2004, \apj, 606, 271

\bibitem[{{Gelman} {et~al}\mbox{.}(2004){Gelman}, {Carlin}, {Stern}, \&
  {Rubin}}]{Gelman+04}
{Gelman} A., {Carlin} J.~B., {Stern} H.~S., {Rubin} D.~B., 2004, {Bayesian Data
  Analysis: Second Edition}. Chapman \& Hall

\bibitem[{{Gelman} \& {Hill}(2007)}]{Gelman&Hill07}
{Gelman} A., {Hill} J., 2007, {Data Analysis Using Regression and
  Multilevel/Hierarchical Modeling}. Cambridge University Press

\bibitem[{{Gil de Paz} {et~al}\mbox{.}(2007){Gil de Paz}, {Boissier}, {Madore},
  {Seibert}, {Joe}, {Boselli}, {Wyder}, {Thilker}, {Bianchi}, {Rey}, {Rich},
  {Barlow}, {Conrow}, {Forster}, {Friedman}, {Martin}, {Morrissey}, {Neff},
  {Schiminovich}, {Small}, {Donas}, {Heckman}, {Lee}, {Milliard}, {Szalay}, \&
  {Yi}}]{GildePaz+07}
{Gil de Paz} A. {et~al.}, 2007, \apjs, 173, 185

\bibitem[{{Glover} \& {Clark}(2012)}]{Glover&Clark12}
{Glover} S.~C.~O., {Clark} P.~C., 2012, \mnras, 421, 9

\bibitem[{{Glover} \& {Mac Low}(2011)}]{Glover&Maclow11}
{Glover} S.~C.~O., {Mac Low} M., 2011, \mnras, 412, 337

\bibitem[{{Heiderman} {et~al}\mbox{.}(2010){Heiderman}, {Evans}, {Allen},
  {Huard}, \& {Heyer}}]{Heiderman+10}
{Heiderman} A., {Evans}, II N.~J., {Allen} L.~E., {Huard} T., {Heyer} M., 2010,
  \apj, 723, 1019

\bibitem[{{Helfer} {et~al}\mbox{.}(2003){Helfer}, {Thornley}, {Regan}, {Wong},
  {Sheth}, {Vogel}, {Blitz}, \& {Bock}}]{Helfer+03}
{Helfer} T.~T., {Thornley} M.~D., {Regan} M.~W., {Wong} T., {Sheth} K., {Vogel}
  S.~N., {Blitz} L., {Bock} D.~C.-J., 2003, \apjs, 145, 259

\bibitem[{{Hopkins} {et~al}\mbox{.}(2011){Hopkins}, {Quataert}, \&
  {Murray}}]{Hopkins+11}
{Hopkins} P.~F., {Quataert} E., {Murray} N., 2011, \mnras, 417, 950

\bibitem[{{Hughes} {et~al}\mbox{.}(2013){Hughes}, {Meidt}, {Schinnerer},
  {Colombo}, {Pety}, {Leroy}, {Dobbs}, {Garc{\'{\i}}a-Burillo}, {Thompson},
  {Dumas}, {Schuster}, \& {Kramer}}]{Hughes+13}
{Hughes} A. {et~al.}, 2013, \apj, 779, 44

\bibitem[{{Kelly}(2007)}]{Kelly07}
{Kelly} B.~C., 2007, \apj, 665, 1489

\bibitem[{{Kennicutt} \& {Evans}(2012)}]{Kennicutt&Evans12}
{Kennicutt} R.~C., {Evans} N.~J., 2012, \araa, 50, 531

\bibitem[{{Kennicutt}(1989)}]{Kennicutt89}
{Kennicutt}, Jr. R.~C., 1989, \apj, 344, 685

\bibitem[{{Kennicutt}(1998)}]{Kennicutt98}
{Kennicutt}, Jr. R.~C., 1998, \apj, 498, 541

\bibitem[{{Kennicutt} {et~al}\mbox{.}(2003){Kennicutt}, {Armus}, {Bendo},
  {Calzetti}, {Dale}, {Draine}, {Engelbracht}, {Gordon}, {Grauer}, {Helou},
  {Hollenbach}, {Jarrett}, {Kewley}, {Leitherer}, {Li}, {Malhotra}, {Regan},
  {Rieke}, {Rieke}, {Roussel}, {Smith}, {Thornley}, \& {Walter}}]{Kennicutt+03}
{Kennicutt}, Jr. R.~C. {et~al.}, 2003, \pasp, 115, 928

\bibitem[{{Kim} {et~al}\mbox{.}(2011){Kim}, {Kim}, \& {Ostriker}}]{Kim+11}
{Kim} C.-G., {Kim} W.-T., {Ostriker} E.~C., 2011, \apj, 743, 25

\bibitem[{{Kim} {et~al}\mbox{.}(2013){Kim}, {Krumholz}, {Wise}, {Turk},
  {Goldbaum}, \& {Abel}}]{Kim+13}
{Kim} J.-h., {Krumholz} M.~R., {Wise} J.~H., {Turk} M.~J., {Goldbaum} N.~J.,
  {Abel} T., 2013, \apj, 779, 8

\bibitem[{{Kruijssen} \& {Longmore}(2014)}]{Kruijssen&Longmore14}
{Kruijssen} J.~M.~D., {Longmore} S.~N., 2014, ArXiv e-prints

\bibitem[{{Krumholz}(2012)}]{Krumholz12}
{Krumholz} M.~R., 2012, \apj, 759, 9

\bibitem[{{Krumholz} {et~al}\mbox{.}(2011){Krumholz}, {Leroy}, \&
  {McKee}}]{Krumholz+11}
{Krumholz} M.~R., {Leroy} A.~K., {McKee} C.~F., 2011, \apj, 731, 25

\bibitem[{{Krumholz} \& {McKee}(2005)}]{Krumholz&McKee05}
{Krumholz} M.~R., {McKee} C.~F., 2005, \apj, 630, 250

\bibitem[{{Krumholz} \& {Thompson}(2007)}]{Krumholz&Thompson07}
{Krumholz} M.~R., {Thompson} T.~A., 2007, \apj, 669, 289

\bibitem[{{Kruschke}(2011)}]{Kruschke11}
{Kruschke} J.~K., 2011, {Doing Bayesian Data Analysis}. Elsevier Inc.

\bibitem[{{Lada} {et~al}\mbox{.}(2012){Lada}, {Forbrich}, {Lombardi}, \&
  {Alves}}]{Lada+12}
{Lada} C.~J., {Forbrich} J., {Lombardi} M., {Alves} J.~F., 2012, \apj, 745, 190

\bibitem[{{Lada} {et~al}\mbox{.}(2010){Lada}, {Lombardi}, \& {Alves}}]{Lada+10}
{Lada} C.~J., {Lombardi} M., {Alves} J.~F., 2010, \apj, 724, 687

\bibitem[{{Lee} {et~al}\mbox{.}(2014){Lee}, {Stanimirovic}, {Wolfire},
  {Shetty}, {Glover}, {Molina}, \& {Klessen}}]{Lee+14}
{Lee} M.-Y., {Stanimirovic} S., {Wolfire} M.~G., {Shetty} R., {Glover}
  S.~C.~O., {Molina} F.~Z., {Klessen} R.~S., 2014, ArXiv e-prints

\bibitem[{{Leroy} {et~al}\mbox{.}(2011){Leroy}, {Bolatto}, {Gordon},
  {Sandstrom}, {Gratier}, {Rosolowsky}, {Engelbracht}, {Mizuno}, {Corbelli},
  {Fukui}, \& {Kawamura}}]{Leroy+11}
{Leroy} A.~K. {et~al.}, 2011, \apj, 737, 12

\bibitem[{{Leroy} {et~al}\mbox{.}(2009){Leroy}, {Walter}, {Bigiel}, {Usero},
  {Weiss}, {Brinks}, {de Blok}, {Kennicutt}, {Schuster}, {Kramer},
  {Wiesemeyer}, \& {Roussel}}]{Leroy+09a}
{Leroy} A.~K. {et~al.}, 2009, \aj, 137, 4670

\bibitem[{{Leroy} {et~al}\mbox{.}(2008){Leroy}, {Walter}, {Brinks}, {Bigiel},
  {de Blok}, {Madore}, \& {Thornley}}]{Leroy+08}
{Leroy} A.~K., {Walter} F., {Brinks} E., {Bigiel} F., {de Blok} W.~J.~G.,
  {Madore} B., {Thornley} M.~D., 2008, \aj, 136, 2782

\bibitem[{{Leroy} {et~al}\mbox{.}(2013){Leroy}, {Walter}, {Sandstrom},
  {Schruba}, {Munoz-Mateos}, {Bigiel}, {Bolatto}, {Brinks}, {de Blok}, {Meidt},
  {Rix}, {Rosolowsky}, {Schinnerer}, {Schuster}, \& {Usero}}]{Leroy+13}
{Leroy} A.~K. {et~al.}, 2013, \aj, 146, 19

\bibitem[{{Liszt} {et~al}\mbox{.}(2010){Liszt}, {Pety}, \& {Lucas}}]{Liszt+10}
{Liszt} H.~S., {Pety} J., {Lucas} R., 2010, \aap, 518, A45+

\bibitem[{{Liu} {et~al}\mbox{.}(2011){Liu}, {Koda}, {Calzetti}, {Fukuhara}, \&
  {Momose}}]{Liu+11}
{Liu} G., {Koda} J., {Calzetti} D., {Fukuhara} M., {Momose} R., 2011, \apj,
  735, 63

\bibitem[{{Mac Low} \& {Klessen}(2004)}]{Maclow&Klessen04}
{Mac Low} M., {Klessen} R.~S., 2004, Reviews of Modern Physics, 76, 125

\bibitem[{{Magnani} {et~al}\mbox{.}(2000){Magnani}, {Hartmann}, {Holcomb},
  {Smith}, \& {Thaddeus}}]{Magnani+00}
{Magnani} L., {Hartmann} D., {Holcomb} S.~L., {Smith} L.~E., {Thaddeus} P.,
  2000, \apj, 535, 167

\bibitem[{{Maloney}(1990)}]{Maloney90}
{Maloney} P., 1990, \apjl, 348, L9

\bibitem[{{Maloney} \& {Black}(1988)}]{Maloney&Black88}
{Maloney} P., {Black} J.~H., 1988, \apj, 325, 389

\bibitem[{{McKee} \& {Ostriker}(2007)}]{McKee&Ostriker07}
{McKee} C.~F., {Ostriker} E.~C., 2007, \araa, 45, 565

\bibitem[{{Momose} {et~al}\mbox{.}(2013){Momose}, {Koda}, {Kennicutt}, {Egusa},
  {Calzetti}, {Liu}, {Donovan Meyer}, {Okumura}, {Scoville}, {Sawada}, \&
  {Kuno}}]{Momose+13}
{Momose} R. {et~al.}, 2013, \apjl, 772, L13

\bibitem[{{Murray}(2011)}]{Murray11}
{Murray} N., 2011, \apj, 729, 133

\bibitem[{{Murray} \& {Rahman}(2010)}]{MurrayRahman10}
{Murray} N., {Rahman} M., 2010, \apj, 709, 424

\bibitem[{{Narayanan} {et~al}\mbox{.}(2008){Narayanan}, {Cox}, {Shirley},
  {Dav{\'e}}, {Hernquist}, \& {Walker}}]{Narayanan+08}
{Narayanan} D., {Cox} T.~J., {Shirley} Y., {Dav{\'e}} R., {Hernquist} L.,
  {Walker} C.~K., 2008, \apj, 684, 996

\bibitem[{{Narayanan} \& {Hopkins}(2013)}]{Narayanan&Hopkins13}
{Narayanan} D., {Hopkins} P.~F., 2013, \mnras, 433, 1223

\bibitem[{{Narayanan} {et~al}\mbox{.}(2011){Narayanan}, {Krumholz}, {Ostriker},
  \& {Hernquist}}]{Narayanan+11}
{Narayanan} D., {Krumholz} M., {Ostriker} E.~C., {Hernquist} L., 2011, \mnras,
  418, 664

\bibitem[{{Narayanan} {et~al}\mbox{.}(2012){Narayanan}, {Krumholz}, {Ostriker},
  \& {Hernquist}}]{Narayanan+12}
{Narayanan} D., {Krumholz} M.~R., {Ostriker} E.~C., {Hernquist} L., 2012,
  \mnras, 421, 3127

\bibitem[{{Onishi} {et~al}\mbox{.}(2001){Onishi}, {Yoshikawa}, {Yamamoto},
  {Kawamura}, {Mizuno}, \& {Fukui}}]{Onishi+01}
{Onishi} T., {Yoshikawa} N., {Yamamoto} H., {Kawamura} A., {Mizuno} A., {Fukui}
  Y., 2001, \pasj, 53, 1017

\bibitem[{{Onodera} {et~al}\mbox{.}(2010){Onodera}, {Kuno}, {Tosaki}, {Kohno},
  {Nakanishi}, {Sawada}, {Muraoka}, {Komugi}, {Miura}, {Kaneko}, {Hirota}, \&
  {Kawabe}}]{Onodera+10}
{Onodera} S. {et~al.}, 2010, \apjl, 722, L127

\bibitem[{{Ostriker} {et~al}\mbox{.}(2010){Ostriker}, {McKee}, \&
  {Leroy}}]{Ostriker+10}
{Ostriker} E.~C., {McKee} C.~F., {Leroy} A.~K., 2010, \apj, 721, 975

\bibitem[{{Ostriker} \& {Shetty}(2011)}]{Ostriker&Shetty11}
{Ostriker} E.~C., {Shetty} R., 2011, \apj, 731, 41

\bibitem[{{Pety} {et~al}\mbox{.}(2013){Pety}, {Schinnerer}, {Leroy}, {Hughes},
  {Meidt}, {Colombo}, {Dumas}, {Garc{\'{\i}}a-Burillo}, {Schuster}, {Kramer},
  {Dobbs}, \& {Thompson}}]{Pety+13}
{Pety} J. {et~al.}, 2013, \apj, 779, 43

\bibitem[{{Polk} {et~al}\mbox{.}(1988){Polk}, {Knapp}, {Stark}, \&
  {Wilson}}]{Polk+88}
{Polk} K.~S., {Knapp} G.~R., {Stark} A.~A., {Wilson} R.~W., 1988, \apj, 332,
  432

\bibitem[{{Pringle} {et~al}\mbox{.}(2001){Pringle}, {Allen}, \&
  {Lubow}}]{Pringle+01}
{Pringle} J.~E., {Allen} R.~J., {Lubow} S.~H., 2001, \mnras, 327, 663

\bibitem[{{Rahman} {et~al}\mbox{.}(2011){Rahman}, {Bolatto}, {Wong}, {Leroy},
  {Walter}, {Rosolowsky}, {West}, {Bigiel}, {Ott}, {Xue}, {Herrera-Camus},
  {Jameson}, {Blitz}, \& {Vogel}}]{Rahman+11}
{Rahman} N. {et~al.}, 2011, \apj, 730, 72

\bibitem[{{Rahman} {et~al}\mbox{.}(2012){Rahman}, {Bolatto}, {Xue}, {Wong},
  {Leroy}, {Walter}, {Bigiel}, {Rosolowsky}, {Fisher}, {Vogel}, {Blitz},
  {West}, \& {Ott}}]{Rahman+12}
{Rahman} N. {et~al.}, 2012, \apj, 745, 183

\bibitem[{{Rosolowsky} {et~al}\mbox{.}(2007){Rosolowsky}, {Keto}, {Matsushita},
  \& {Willner}}]{Rosolowsky+07}
{Rosolowsky} E., {Keto} E., {Matsushita} S., {Willner} S.~P., 2007, \apj, 661,
  830

\bibitem[{{Rubio} {et~al}\mbox{.}(2004){Rubio}, {Boulanger}, {Rantakyro}, \&
  {Contursi}}]{Rubio+04}
{Rubio} M., {Boulanger} F., {Rantakyro} F., {Contursi} A., 2004, \aap, 425, L1

\bibitem[{{Sandstrom} {et~al}\mbox{.}(2013){Sandstrom}, {Leroy}, {Walter},
  {Bolatto}, {Croxall}, {Draine}, {Wilson}, {Wolfire}, {Calzetti}, {Kennicutt},
  {Aniano}, {Donovan Meyer}, {Usero}, {Bigiel}, {Brinks}, {de Blok}, {Crocker},
  {Dale}, {Engelbracht}, {Galametz}, {Groves}, {Hunt}, {Koda}, {Kreckel},
  {Linz}, {Meidt}, {Pellegrini}, {Rix}, {Roussel}, {Schinnerer}, {Schruba},
  {Schuster}, {Skibba}, {van der Laan}, {Appleton}, {Armus}, {Brandl},
  {Gordon}, {Hinz}, {Krause}, {Montiel}, {Sauvage}, {Schmiedeke}, {Smith}, \&
  {Vigroux}}]{Sandstrom+13}
{Sandstrom} K.~M. {et~al.}, 2013, \apj, 777, 5

\bibitem[{{Scalo}(1990)}]{Scalo90}
{Scalo} J., 1990, in Astrophysics and Space Science Library, Vol. 162, Physical
  Processes in Fragmentation and Star Formation, {Capuzzo-Dolcetta} R.,
  {Chiosi} C., {di Fazio} A., eds., pp. 151--176

\bibitem[{{Scalo} \& {Elmegreen}(2004)}]{Scalo&Elmegreen04}
{Scalo} J., {Elmegreen} B.~G., 2004, \araa, 42, 275

\bibitem[{{Schinnerer} {et~al}\mbox{.}(2013){Schinnerer}, {Meidt}, {Pety},
  {Hughes}, {Colombo}, {Garc{\'{\i}}a-Burillo}, {Schuster}, {Dumas}, {Dobbs},
  {Leroy}, {Kramer}, {Thompson}, \& {Regan}}]{Schinnerer+13}
{Schinnerer} E. {et~al.}, 2013, \apj, 779, 42

\bibitem[{{Schmidt}(1959)}]{Schmidt59}
{Schmidt} M., 1959, \apj, 129, 243

\bibitem[{{Schruba} {et~al}\mbox{.}(2011){Schruba}, {Leroy}, {Walter},
  {Bigiel}, {Brinks}, {de Blok}, {Dumas}, {Kramer}, {Rosolowsky}, {Sandstrom},
  {Schuster}, {Usero}, {Weiss}, \& {Wiesemeyer}}]{Schruba+11}
{Schruba} A. {et~al.}, 2011, \aj, 142, 37

\bibitem[{{Schruba} {et~al}\mbox{.}(2010){Schruba}, {Leroy}, {Walter},
  {Sandstrom}, \& {Rosolowsky}}]{Schruba+10}
{Schruba} A., {Leroy} A.~K., {Walter} F., {Sandstrom} K., {Rosolowsky} E.,
  2010, \apj, 722, 1699

\bibitem[{{Shetty} {et~al}\mbox{.}(2010){Shetty}, {Collins}, {Kauffmann},
  {Goodman}, {Rosolowsky}, \& {Norman}}]{Shetty+10}
{Shetty} R., {Collins} D.~C., {Kauffmann} J., {Goodman} A.~A., {Rosolowsky}
  E.~W., {Norman} M.~L., 2010, \apj, 712, 1049

\bibitem[{{Shetty} {et~al}\mbox{.}(2011{\natexlab{a}}){Shetty}, {Glover},
  {Dullemond}, \& {Klessen}}]{Shetty+11a}
{Shetty} R., {Glover} S.~C., {Dullemond} C.~P., {Klessen} R.~S.,
  2011{\natexlab{a}}, \mnras, 412, 1686

\bibitem[{{Shetty} {et~al}\mbox{.}(2011{\natexlab{b}}){Shetty}, {Glover},
  {Dullemond}, {Ostriker}, {Harris}, \& {Klessen}}]{Shetty+11b}
{Shetty} R., {Glover} S.~C., {Dullemond} C.~P., {Ostriker} E.~C., {Harris}
  A.~I., {Klessen} R.~S., 2011{\natexlab{b}}, \mnras, 415, 3253

\bibitem[{{Shetty} {et~al}\mbox{.}(2013){Shetty}, {Kelly}, \&
  {Bigiel}}]{Shetty+13}
{Shetty} R., {Kelly} B.~C., {Bigiel} F., 2013, \mnras, 430, 288

\bibitem[{{Shetty} {et~al}\mbox{.}(2014){Shetty}, {Kelly}, {Rahman}, {Bigiel},
  {Bolatto}, {Clark}, {Klessen}, \& {Konstandin}}]{Shetty+14}
{Shetty} R., {Kelly} B.~C., {Rahman} N., {Bigiel} F., {Bolatto} A.~D., {Clark}
  P.~C., {Klessen} R.~S., {Konstandin} L.~K., 2014, \mnras, 437, L61

\bibitem[{{Shetty} \& {Ostriker}(2008)}]{Shetty&Ostriker08}
{Shetty} R., {Ostriker} E.~C., 2008, \apj, 684, 978

\bibitem[{{Solomon} \& {Rivolo}(1989)}]{SolomonRivolo89}
{Solomon} P.~M., {Rivolo} A.~R., 1989, \apj, 339, 919

\bibitem[{{Solomon} {et~al}\mbox{.}(1987){Solomon}, {Rivolo}, {Barrett}, \&
  {Yahil}}]{Solomon+87}
{Solomon} P.~M., {Rivolo} A.~R., {Barrett} J., {Yahil} A., 1987, \apj, 319, 730

\bibitem[{{van Dishoeck} \& {Black}(1988)}]{vanDishoeck&Black88}
{van Dishoeck} E.~F., {Black} J.~H., 1988, \apj, 334, 771

\bibitem[{{Wall}(2007)}]{Wall07}
{Wall} W.~F., 2007, \mnras, 379, 674

\bibitem[{{Wilson} \& {Walker}(1994)}]{Wilson&Walker94}
{Wilson} C.~D., {Walker} C.~E., 1994, \apj, 432, 148

\bibitem[{{Wilson} {et~al}\mbox{.}(2012){Wilson}, {Warren}, {Israel},
  {Serjeant}, {Attewell}, {Bendo}, {Butner}, {Chanial}, {Clements}, {Golding},
  {Heesen}, {Irwin}, {Leech}, {Matthews}, {M{\"u}hle}, {Mortier}, {Petitpas},
  {S{\'a}nchez-Gallego}, {Sinukoff}, {Shorten}, {Tan}, {Tilanus}, {Usero},
  {Vaccari}, {Wiegert}, {Zhu}, {Alexander}, {Alexander}, {Azimlu}, {Barmby},
  {Brar}, {Bridge}, {Brinks}, {Brooks}, {Coppin}, {C{\^o}t{\'e}},
  {C{\^o}t{\'e}}, {Courteau}, {Davies}, {Eales}, {Fich}, {Hudson}, {Hughes},
  {Ivison}, {Knapen}, {Page}, {Parkin}, {Rigopoulou}, {Rosolowsky}, {Seaquist},
  {Spekkens}, {Tanvir}, {van der Hulst}, {van der Werf}, {Vlahakis}, {Webb},
  {Weferling}, \& {White}}]{Wilson+12}
{Wilson} C.~D. {et~al.}, 2012, \mnras, 424, 3050

\bibitem[{{Wilson} {et~al}\mbox{.}(2009){Wilson}, {Warren}, {Israel},
  {Serjeant}, {Bendo}, {Brinks}, {Clements}, {Courteau}, {Irwin}, {Knapen},
  {Leech}, {Matthews}, {M{\"u}hle}, {Mortier}, {Petitpas}, {Sinukoff},
  {Spekkens}, {Tan}, {Tilanus}, {Usero}, {van der Werf}, {Wiegert}, \&
  {Zhu}}]{Wilson+09}
{Wilson} C.~D. {et~al.}, 2009, \apj, 693, 1736

\bibitem[{{Wu} {et~al}\mbox{.}(2005){Wu}, {Evans}, {Gao}, {Solomon}, {Shirley},
  \& {Vanden Bout}}]{Wu+05}
{Wu} J., {Evans}, II N.~J., {Gao} Y., {Solomon} P.~M., {Shirley} Y.~L., {Vanden
  Bout} P.~A., 2005, \apjl, 635, L173

\end{thebibliography}
\bibliographystyle{mn2e}

\label{lastpage}

\end{document}